\documentclass{ws-ijmpa1}
\usepackage{graphicx}
\usepackage{bm}
\usepackage{psfrag}
\usepackage{amsmath}
\usepackage{amssymb}
\usepackage{latexsym}
\usepackage{exscale}

\newcommand{\vect}[1]{\bm{#1}}
\newcommand{\ten}[1]{\mbox{\textbf{{\textsf{#1}}}}}

\newcommand{\veczero}{\mbox{\textbf{\textit{0}}}}

\newcommand{\sprod}{\!\cdot\!}
\newcommand{\tprod}{}
\newcommand{\vprod}{\!\times\!}
\newcommand{\trace}{\operatorname{Tr}}
\newcommand{\trans}{{\operatorname{T}}}
\newcommand{\dif}{\mathrm{d}}
\newcommand{\mi}{\mathrm{i}} 

\begin{document}
\markboth{S.~Y.~Buhmann, S.~Scheel, H.~Safari and D.-G.~Welsch}
{Dispersion forces and duality}
%
%
%
%
\title{DISPERSION FORCES AND DUALITY}
\author{STEFAN YOSHI BUHMANN, STEFAN SCHEEL}
\address{Quantum Optics and Laser Science, Blackett Laboratory,
Imperial College London, Prince Consort Road,
London SW7 2AZ, United Kingdom\\
s.buhmann@imperial.ac.uk}
\author{HASSAN SAFARI, DIRK-GUNNAR WELSCH}
\address{Theoretisch-Physikalisches Institut,
Friedrich-Schiller-Universit\"{a}t Jena, Max-Wien-Platz 1, 07743 Jena,
Germany}
\maketitle
\begin{history}
\received{\today}
\end{history}
\begin{abstract}
We formulate a symmetry principle on the basis of the duality of
electric and magnetic fields and apply it to dispersion forces. Within
the context of macroscopic quantum electrodynamics, we rigorously
establish duality invariance for the free electromagnetic field in the
presence of causal magnetoelectrics. Dispersion forces are given in
terms of the Green tensor for the electromagnetic field and the atomic
response functions. After discussing the behavior of the Green tensor
under a duality transformation, we are able to show that Casimir
forces on bodies in free space as well as local-field corrected
Casimir--Polder and van der Waals forces are duality invariant.
\end{abstract}
\ccode{PACS numbers: 
12.20.--m, 
34.35.+a,  
34.20.--b, 
42.50.Nn   
}
%
%
\section{Introduction}	
Dispersion forces such as the Casimir force\cite{0373} on a body, the
Casimir--Polder (CP) force\cite{0030} between an atom and a body and
the van der Waals (vdW) force\cite{0030} between two atoms were
originally conceived as effective electromagnetic forces between
electrically neutral, but polarizable ground-state objects; they are
typically attractive.\cite{0696} Somewhat later, the investigations 
were extended to bodies and atoms with an additional magnetic
response,\cite{0089} revealing that polarizable and magnetizable
objects may repel each other. Forces between magnetoelectric systems
have recently been subject to a renewed interest\cite{0121} due to the
availability of metamaterials with a controllable permittivity and
permeability.\cite{0479}

As we will demonstrate in this article, investigations of this kind
can be considerably simplified by exploiting the well-known duality of
electric and magnetic fields.\cite{0845} To that end, we first study
the effect of duality transformations in the context of macroscopic
quantum electrodynamics\cite{0696} (Sec.~\ref{sec2}) and then use our
results to prove that duality invariance is a valid symmetry of
dispersion forces under very general conditions (Sec.~\ref{sec3}).
%
%
\section{Macroscopic quantum electrodynamics and duality}	
\label{sec2}
Duality is one of the inherent symmetries of the Maxwell equations.
To see this, we group the electric and magnetic fields
$\hat{\vect{E}}(\vect{r},t)$, $\hat{\vect{B}}(\vect{r},t)$ and
excitations $\hat{\vect{D}}(\vect{r},t)$, $\hat{\vect{H}}(\vect{r},t)$
into dual pairs $(\hat{\vect{E}},Z_0\hat{\vect{H}})^\trans$ and
$(Z_0\hat{\vect{D}},\hat{\vect{B}})^\trans$, where the vacuum
impedance $Z_0\!=\!\sqrt{\mu_0/\varepsilon_0}$ has been introduced for
dimensional reasons. In this dual-pair notation, the Maxwell equations
in the absence of free charges and currents assume the compact form
\begin{gather}
\label{eq1}
\vect{\nabla}\sprod
 \biggl(\begin{array}{c}Z_0\hat{\vect{D}}\\
 \hat{\vect{B}}\end{array}\biggr)
=\biggl(\begin{array}{c}0\\ 0\end{array}\biggr),\\
\label{eq2}
\vect{\nabla}\vprod
 \biggl(\begin{array}{c}\hat{\vect{E}}\\ 
 Z_0\hat{\vect{H}}\end{array}\biggr)
 +\frac{\partial}{\partial t}
 \biggl(\begin{array}{cc}0&1\\-1&0\end{array}\biggr)
 \biggl(\begin{array}{c}Z_0\hat{\vect{D}}\\
 \hat{\vect{B}}\end{array}\biggr)
 =\biggl(\begin{array}{c}\veczero\\ \veczero\end{array}\biggr).
\end{gather}
Grouping the polarization $\hat{\vect{P}}(\vect{r},t)$ and the
magnetization $\hat{\vect{M}}(\vect{r},t)$ according to
$(Z_0\hat{\vect{P}},\mu_0\hat{\vect{M}})^\trans$, 
the relation between the fields and excitations reads
\begin{equation}
\label{eq3}
 \biggl(\begin{array}{c}Z_0\hat{\vect{D}}\\
\hat{\vect{B}}\end{array}\biggr)
 =\frac{1}{c}
 \biggl(\begin{array}{c}\hat{\vect{E}}\\
 Z_0\hat{\vect{H}}\end{array}\biggr)
 +\biggl(\begin{array}{c}Z_0\hat{\vect{P}}\\
 \mu_0\hat{\vect{M}}\end{array}\biggr).
\end{equation}
It is now immediately obvious that Eqs.~(\ref{eq1})--(\ref{eq3}) are
invariant with respect to a duality transformation
\begin{equation}
\label{eq4}
\biggl(\begin{array}{c}\vect{x}\\ \vect{y}\end{array}\biggr)^\star
 =\mathcal{D}(r,\theta)
 \biggl(\begin{array}{c}\vect{x}\\ \vect{y}\end{array}\biggr),
 \qquad\mathcal{D}(r,\theta)
 =\biggl(\begin{array}{cc}r\cos\theta&r\sin\theta\\ 
 -r\sin\theta&r\cos\theta\end{array}\biggr)\in\mathbb{R}_+\!\times
 \operatorname{SO}(2),
\end{equation}
with $\mathcal{D}(r,\theta)$ being the most general real matrix that 
commutes with the symplectic matrix in Eq.~(\ref{eq2}). This
transformation may be viewed as a rotation in the space of dual pairs
($0\!\le\!\theta\!<\!2\pi$) together with a rescaling of all fields
($r\!>\!0$). In classical physics, the real-valued electromagnetic
fields are often combined into complex Riemann--Silberstein vectors
$\vect{x}\!+\!\mi\vect{y}$,\cite{0844} in which case duality
invariance manifests itself as a $\operatorname{U}(1)$ symmetry. 

Let us next address the compatibility of the duality transformation
with the constitutive relations, which may conveniently be formulated
in terms of the Fourier components $\underline{\hat{\vect{x}}}$ of the
fields,
$\hat{\vect{x}}(\vect{r},t)\!=\!\int_0^\infty\!\dif\omega\,
\underline{\hat{\vect{x}}}(\vect{r},\omega,t)+$h.c. For linear,
local, isotropic, dispersing and absorbing media, the
constitutive relations may be given as
\begin{equation}
\label{eq5}
 \biggl(\begin{array}{c}Z_0\hat{\underline{\vect{D}}}\\
 \hat{\underline{\vect{B}}}\end{array}\biggr)
 =\frac{1}{c}\biggl(\begin{array}{cc}\varepsilon&0\\
 0&\mu\end{array}\biggr)
 \biggl(\begin{array}{c}\hat{\underline{\vect{E}}}\\
 Z_0\hat{\underline{\vect{H}}}\end{array}\biggr)
 +\biggl(\begin{array}{cc}1&0\\0&\mu\end{array}\biggr)
 \biggl(\begin{array}{c}Z_0\hat{\underline{\vect{P}}}_\mathrm{N}\\
 \mu_0\hat{\underline{\vect{M}}}_\mathrm{N}\end{array}\biggr),
\end{equation}
where $\varepsilon=\varepsilon(\vect{r},\omega)$ and
$\mu=\mu(\vect{r},\omega)$ denote the relative electric permittivity
and magnetic permeability; and $\hat{\vect{P}}_\mathrm{N}$ and
$\hat{\vect{M}}_\mathrm{N}$ are the noise polarization and
magnetization which necessarily arise in the presence of absorbing
media. Invariance of the constitutive relations under the duality
transformation requires that 
\begin{gather}
\label{eq6}
\biggl(\begin{array}{cc}\varepsilon^\star&0\\
 0&\mu^\star\end{array}\biggr)
=\mathcal{D}(r,\theta)
\biggl(\begin{array}{cc}\varepsilon&0\\0&\mu\end{array}\biggr)
\mathcal{D}^{-1}(r,\theta)
=\biggl(\begin{array}{cc}\varepsilon\cos^2\theta+\mu\sin^2\theta
 &(\mu-\varepsilon)\sin\theta\cos\theta\\
 (\mu-\varepsilon)\sin\theta\cos\theta
 &\varepsilon\sin^2\theta+\mu\cos^2\theta\end{array}\biggr),\\
\label{eq7}
\biggl(\begin{array}{c}\hat{\underline{\vect{P}}}_\mathrm{N}\\
 \hat{\underline{\vect{M}}}_\mathrm{N}/c
 \end{array}\biggr)^\star
=\biggl(\begin{array}{cc}r\cos\theta&\mu r\sin\theta\\
 -(1/\mu^\star)r\sin\theta&(\mu/\mu^\star)r\cos\theta\end{array}
 \biggr)
 \biggl(\begin{array}{c}\hat{\underline{\vect{P}}}_\mathrm{N}\\
 \hat{\underline{\vect{M}}}_\mathrm{N}/c
 \end{array}\biggr).
\end{gather}
Condition~(\ref{eq6}) can be fulfilled in two ways: It holds if the
relative impedance of the media is equal to unity,
$Z\!=\!\sqrt{\mu/\varepsilon}\!=\!1$. In this case, which includes
both free space and a perfect lens medium
($\varepsilon\!=\!\mu\!=\!-1$)\cite{0478}, the duality rotations form
a continuous $\operatorname{SO}(2)$ symmetry of the electromagnetic
field and one has $\varepsilon^\star\!=\!\mu^\star\!=\!\varepsilon$ as
well as
\begin{equation}
\label{eq8}
\biggr(\begin{array}{c}\hat{\underline{\vect{P}}}_\mathrm{N}\\
 \hat{\underline{\vect{M}}}_\mathrm{N}/c
 \end{array}\biggr)^\star
=\biggl(\begin{array}{cc}r\cos\theta&\varepsilon r\sin\theta\\
 -(1/\varepsilon)r\sin\theta&r\cos\theta
 \end{array}\biggr)
 \biggl(\begin{array}{c}\hat{\underline{\vect{P}}}_\mathrm{N}\\
 \hat{\underline{\vect{M}}}_\mathrm{N}/c
 \end{array}\biggr).
\end{equation}
For media with a nontrivial impedance, Eq.~(\ref{eq6}) holds for
$\theta\!=\!n\pi/2$ with $n\!\in\!\mathbb{Z}$ only. The presence
of such media hence reduces the duality invariance from the full
$\operatorname{SO}(2)$ group to a discrete $\mathbb{Z}_4$ symmetry
with the four distinct members
\begin{equation}
\label{eq9}
\mathcal{D}_0=\mathcal{I},\quad 
\mathcal{D}_1=\biggl(\begin{array}{cc}0&1\\-1&0\end{array}\biggr),
\quad\mathcal{D}_2=-\mathcal{I},\quad
\mathcal{D}_3=-\mathcal{D}_1,
\end{equation}
($\mathcal{I}$: unit matrix) where Eqs.~(\ref{eq6}) and (\ref{eq7})
imply the transformations
\begin{gather}
\label{eq10}
 \biggl(\begin{array}{c}\varepsilon\\ \mu\end{array}\biggr)^\star
 =\biggl(\begin{array}{cc}\cos^2\theta&\sin^2\theta\\
 \sin^2\theta&\cos^2\theta\end{array}\biggr)
 \biggl(\begin{array}{c}\varepsilon\\ \mu\end{array}\biggr),\\
\label{eq11}
\biggl(\begin{array}{c}\hat{\underline{\vect{P}}}_\mathrm{N}\\
 \hat{\underline{\vect{M}}}_\mathrm{N}/c\end{array}\biggr)^\star
=\biggl(\begin{array}{cc}r\cos\theta&\mu r\sin\theta\\ 
 -(1/\varepsilon)r\sin\theta&
 r\cos\theta\end{array}\biggr)
\biggl(\begin{array}{c}\hat{\underline{\vect{P}}}_\mathrm{N}\\
 \hat{\underline{\vect{M}}}_\mathrm{N}/c\end{array}\biggr).
\end{gather}

Duality is thus an exact symmetry of the Maxwell equations in the
absence of free charges and currents. It must also be manifest in the
underlying Hamiltonian\cite{0002}
$\hat{H}_F\!=\!\sum_{\lambda=e,m}\int\dif^3r%
\int_0^\infty \dif\omega\,\hbar\omega\,%
\hat{\vect{f}}_\lambda^\dagger(\vect{r},\omega)%
\sprod\hat{\vect{f}}_\lambda(\vect{r},\omega)$, where the bosonic
dynamical variables $\hat{\vect{f}}_e(\vect{r},\omega)$,
$\hat{\vect{f}}_m(\vect{r},\omega)$  are associated with $e$lectric
and $m$agnetic medium--field excitations. Noting that the dynamical
variables are related to the noise fields via
\begin{equation}
\label{eq13}
\biggl(\begin{array}{c}Z_0\hat{\underline{\vect{P}}}_\mathrm{N}\\
 \mu_0\hat{\underline{\vect{M}}}_\mathrm{N}
 \end{array}\biggr)
=\sqrt{\frac{\hbar\mu_0}{\pi}}
\biggl(\begin{array}{cc}\mi\sqrt{\operatorname{Im}\varepsilon}&0\\ 
 0&\sqrt{\operatorname{Im}\mu}/|\mu|\end{array}\biggr)
\biggl(\begin{array}{c}\hat{\vect{f}}_e\\
 \hat{\vect{f}}_m\end{array}\biggr)
\end{equation}
and recalling Eqs.~(\ref{eq8}) and (\ref{eq11}), they are seen to
transform as 
\begin{equation}
\label{eq14}
\biggl(\begin{array}{c}\hat{\vect{f}}_e\\
\hat{\vect{f}}_m\end{array}\biggr)^\star
 =\biggl(\begin{array}{cc}r\cos\theta
 &-\mi(\mu/|\mu|)r\sin\theta\\
 -\mi(|\varepsilon|/\varepsilon)r\sin\theta 
 &r\cos\theta\end{array}\biggr)
 \biggl(\begin{array}{c}\hat{\vect{f}}_e\\
 \hat{\vect{f}}_m\end{array}\biggr)
\end{equation}
in both the continuous ($\varepsilon\!=\!\mu$) and discrete
($\theta\!=\!n\pi/2$) cases. It follows that a duality transformation
leads to a rescaling of the Hamiltonian
$\hat{H}^\star_F\!=\!r^2\hat{H}_F$, such that the equations of motion
remain invariant.

It is important to note that electromagnetic forces are not
duality-invariant in general, even when acting on electrically neutral
systems. For instance, the Lorentz force on a neutral magnetoelectric
body of volume $V$ can be written as\cite{0696}
\begin{multline}
\label{eq15}
\hat{\vect{F}}
=\int_{\partial V}\dif\vect{A}\sprod\biggl\{
 \varepsilon_0\hat{\vect{E}}(\vect{r})
 \tprod\hat{\vect{E}}(\vect{r})
 +\frac{1}{\mu_0}\hat{\vect{B}}(\vect{r})
 \tprod\hat{\vect{B}}(\vect{r})
 -\frac{1}{2}\biggl[\varepsilon_0
 \hat{\vect{E}}^2(\vect{r})
 +\frac{1}{\mu_0}\hat{\vect{B}}^2(\vect{r})\biggr]\ten{I}\\
 -\varepsilon_0\,\frac{\dif}{\dif t}
 \int_V \dif^3r\,\hat{\vect{E}}(\vect{r})
 \vprod\hat{\vect{B}}(\vect{r})
\end{multline}
($\ten{I}$: unit tensor); it is obviously not duality-invariant.
Duality invariance would be realized for a stationary field acting on
a body at rest (such that the total time derivative vanishes),
provided that $\hat{\vect{D}}\!\approx\!\varepsilon_0\hat{\vect{E}}$
and $\hat{\vect{H}}\!\approx\!\hat{\vect{B}}/\mu_0$ on the body
surface. While this can never be true on an operator level due to the
unavoidable presence of the noise fields $\hat{\vect{P}}_\mathrm{N}$
and $\hat{\vect{M}}_\mathrm{N}$, one would expect the Casimir force
$\langle\hat{\vect{F}}\rangle$ on a stationary body in free space to
be invariant. The situation is very similar for the Lorentz force on a
neutral atom (position~$\vect{r}_A$, polarization~$\hat{\vect{P}}_A$,
magnetization~$\hat{\vect{M}}_A$) which can be written
as\cite{0696,0008}
\begin{multline}
\label{eq16}
\hat{\vect{F}}
=\vect{\nabla}_{\!A}\int\dif^3r\,\Bigl[
 \hat{\vect{P}}_A(\vect{r})\sprod\hat{\vect{E}}(\vect{r})
 +\hat{\vect{M}}_A(\vect{r})\sprod\hat{\vect{B}}(\vect{r})
 +\hat{\vect{P}}_A(\vect{r})\vprod\dot{\hat{\vect{r}}}_{A}
 \sprod\hat{\vect{B}}(\vect{r})\Bigr]\\
+\frac{\dif}{\dif t}\int\dif^3r\,
 \hat{\vect{P}}_A(\vect{r})\vprod\hat{\vect{B}}(\vect{r})
\end{multline}
when neglecting diamagnetic interactions. Noting that
$\hat{\vect{P}}_A$ and $\hat{\vect{M}}_A$ transform under duality like
$\hat{\vect{P}}$ and $\hat{\vect{M}}$, one sees that duality
invariance can only hold for an atom at rest (so that the
velocity-dependent terms do not contribute) prepared in an incoherent
superposition of energy eigenstates and subject to a stationary field
(so that the time-derivative does not contribute\cite{0008}),
provided that $\hat{\vect{D}}\!\approx\!\varepsilon_0\hat{\vect{E}}$
and $\hat{\vect{H}}\!\approx\!\hat{\vect{B}}/\mu_0$ hold within the
volume occupied by the atom. Again, this is never true on an operator
level, but one may expect CP and vdW forces on atoms to be duality
invariant. We will confirm the conjectured duality invariance of
dispersion forces in Sec.~\ref{sec3}. In the following, we concentrate
our attention on the particular transformation $r\!=\!1$,
$\theta\!=\!\pi/2$ which is a generator of the discrete duality group
$\mathbb{Z}_4$. 
%
%
\section{Duality invariance of dispersion forces}
\label{sec3}
As a preparation for studying the duality invariance of dispersion
forces, let us first establish a few general expressions for these
forces in terms of the relevant response functions. By taking the
ground-state expectation value of the Lorentz force~(\ref{eq15}), one
easily finds that the Casimir force on a stationary homogeneous
magnetoelectric body is given by\cite{0663}
\begin{multline}
\label{eq17}
\vect{F}
=\frac{\hbar}{\pi}\int_{V}\dif^3r\int_{0}^\infty\dif\omega\,
 \biggl(\frac{\omega^2}{c^2}\bm{\nabla}\sprod
 \mathrm{Im}\ten{G}^{(1)}(\vect{r},\vect{r},\omega)\\
+\trace\biggl\{\ten{I}\vprod
 \biggl[\bm{\nabla}\vprod\bm{\nabla}\vprod\,
 -\frac{\omega^2}{c^2}\biggr]
 \mathrm{Im}\ten{G}^{(1)}(\vect{r},\vect{r},\omega)\vprod
 \overleftarrow{\bm{\nabla}}'\biggr\}
 \biggr)
\end{multline}
where $\ten{G}^{(1)}$ is the scattering part of the Green tensor
$\ten{G}$ of the electromagnetic field,
\begin{equation}
\label{eq18}
\biggl[\vect{\nabla}\vprod
 \,\frac{1}{\mu(\vect{r},\omega)}\,\vect{\nabla}\vprod
 \,-\,\frac{\omega^2}{c^2}\,\varepsilon(\vect{r},\omega)\biggr]
 \ten{G}(\vect{r},\vect{r}',\omega)
 =\bm{\delta}(\vect{r}-\vect{r}').
\end{equation}
Alternatively, the Casimir force may be given as a surface
integral\cite{0198} $\vect{F}\!=\!\vect{F}_e\!+\!\vect{F}_m$,
\begin{equation}
\label{eq20}
\vect{F}_\lambda=\frac{\hbar}{\pi}\int_{0}^{\infty} \dif\xi
 \int_{\partial V}\dif\vect{A}\sprod\Bigl[
 \ten{G}_{\lambda\lambda}^{(1)}(\vect{r},\vect{r},\mi\xi)
 -{\textstyle\frac{1}{2}}\ten{I}\trace
 \ten{G}_{\lambda\lambda}^{(1)}(\vect{r},\vect{r},\mi\xi)\Bigr]
 \quad(\lambda=e,m)
\end{equation}
with $\ten{G}_{ee}(\vect{r},\vect{r}',\omega)%
\!=\!(\mi\omega/c)\ten{G}(\vect{r},\vect{r}',\omega)(\mi\omega/c)$,
$\ten{G}_{mm}(\vect{r},\vect{r}',\omega)%
\!=\!\vect{\nabla}\vprod\ten{G}(\vect{r},\vect{r}',\omega)\vprod
\overleftarrow{\vect{\nabla}}'$.

The CP force on a single atom can be derived from the Casimir force in
its volume-integral form~(\ref{eq17}) by considering the force on a
dilute gas of atoms [number density $\eta(\vect{r})$, polarizability
$\alpha(\omega)$, magnetizability $\beta(\omega)$] occupying an
otherwise empty volume $V$. Within leading order in $\eta$, the
contributions of the atoms to the permittivity and the inverse
permeability ($\kappa\!=\!\mu^{-1}$) is given by the linearized
Clausius--Mosotti relations\cite{0001}
$\Delta\varepsilon(\vect{r},\omega)\!=\!\eta(\vect{r})
\alpha(\omega)/\varepsilon_0$ and
$\Delta\kappa(\vect{r},\omega)\!=\!-\eta(\vect{r})
\beta(\omega)\mu_0$. Using a linear Born expansion\cite{0491}, one
finds that the resulting change of the Green tensor reads
\begin{multline}
\label{eq22}
\Delta\ten{G}(\vect{r},\vect{r}',\omega)
 =\int\dif^3s\,\frac{\eta(\vect{s})}{\varepsilon_0}\biggl\{
 \frac{\omega^2}{c^2}\,\alpha(\omega)
 \ten{G}(\vect{r},\vect{s},\omega)
 \sprod\ten{G}(\vect{s},\vect{r}',\omega)\\
-\frac{\beta(\omega)}{c^2}
 \Bigl[\ten{G}(\vect{r},\vect{s},\omega)\vprod
 \overleftarrow{\vect{\nabla}}_{\!\vect{s}}\Bigr]
 \sprod\Bigl[
 \vect{\nabla}_{\!\vect{s}}\vprod
 \ten{G}(\vect{s},\vect{r}',\omega)\Bigr]\biggr\}.
\end{multline}
Combining this with Eq.~(\ref{eq17}) one can show that the Casimir
force on the atomic cloud can be written as\cite{0663}
$\vect{F}%
=-\int_{V}\dif^3r\,\eta(\vect{r})\vect{\nabla}U(\vect{r})$ where
$U(\vect{r}_A)\!=\!U_e(\vect{r}_A)\!+\!U_e(\vect{r}_A)$ with
\begin{equation}
\label{eq23}
U_\lambda(\vect{r}_A)
 =\frac{\hbar}{2\pi\varepsilon_0}
 \int_0^\infty\dif\xi\,\alpha_\lambda(\mi\xi)
 \trace\ten{G}_{\lambda\lambda}^{(1)}(\vect{r}_A,\vect{r}_A,\mi\xi)
 \quad(\lambda=e,m)
\end{equation}
($\alpha_e\!=\!\alpha$, $\alpha_m=\beta/c^2$) is the CP potential
sought. The vdW potential between two atoms $A$ and $B$ can be
obtained in an analogous way by introducing a second dilute cloud of
atoms and applying Eq.~(\ref{eq22}) to the CP potential. This results
in $U(\vect{r}_A)%
 =\int_{V}\dif^3r\,\eta(\vect{r})U(\vect{r}_A,\vect{r})$, where the
required two-atom vdW potential reads $U(\vect{r}_A,\vect{r}_B)%
\!=\!U_{ee}(\vect{r}_A,\vect{r}_B)\!+\!U_{em}(\vect{r}_A,\vect{r}_B)%
\!+\!U_{me}(\vect{r}_A,\vect{r}_B)\!+\!U_{mm}(\vect{r}_A,\vect{r}_B)$,
\begin{multline}
\label{eq26}
U_{\lambda\lambda'}(\vect{r}_A,\vect{r}_B)
 =-\frac{\hbar}{2\pi\varepsilon_0^2}\int_0^\infty\!\!\dif\xi\,
\alpha_\lambda^A(\mi\xi)\alpha_{\lambda'}^B(\mi\xi) 
 \trace\Bigl[\ten{G}_{\lambda\lambda'}(\vect{r}_A,\vect{r}_B,\mi\xi)
 \sprod\ten{G}_{\lambda'\lambda}(\vect{r}_B,\vect{r}_{\!A},\mi\xi)
 \Bigr]\\
(\lambda,\lambda'=e,m)
\end{multline}
with $\ten{G}_{em}(\vect{r},\vect{r}',\omega)%
\!=\!(\mi\omega/c)\ten{G}(\vect{r},\vect{r}',\omega)\vprod%
\overleftarrow{\vect{\nabla}}'$,
$\ten{G}_{me}(\vect{r},\vect{r}',\omega)%
\!=\!\vect{\nabla}\vprod\ten{G}(\vect{r},\vect{r}',\omega)%
(\mi\omega/c)$.

Dispersion forces and potentials can thus be given in terms of the
response functions of the electromagnetic field and the atoms, so
their behavior under a duality transformation can be determined from
that of $\ten{G}$, $\alpha$ and $\beta$. By virtue of the linearized
Clausius--Mosotti relations, the known transformation properties
$\varepsilon^\star\!=\!\mu$, $\mu^\star\!=\!\varepsilon$ imply that
$\alpha^\star=c^2\beta$, $\beta^\star=\alpha/c^2$. As shown in
\ref{appA}, the Green tensor transforms according to 
\begin{eqnarray}
\label{eq27}
\ten{G}_{ee}^\star(\vect{r},\vect{r}',\omega)
&=&\mu^{-1}(\vect{r},\omega)
 \ten{G}_{mm}(\vect{r},\vect{r}',\omega)
 \mu^{-1}(\vect{r}',\omega)
+\mu^{-1}(\vect{r},\omega)
 \bm{\delta}(\vect{r}\!-\!\vect{r}'),\\
\label{eq28}
\ten{G}_{mm}^\star(\vect{r},\vect{r}',\omega)
&=&\varepsilon(\vect{r},\omega)
 \ten{G}_{ee}(\vect{r},\vect{r}',\omega)\,
 \varepsilon(\vect{r}',\omega)
 -\varepsilon(\vect{r},\omega)\bm{\delta}(\vect{r}\!-\!\vect{r}'),\\
\label{eq29}
\ten{G}_{em}^\star(\vect{r},\vect{r}',\omega)
&=&-\mu^{-1}(\vect{r},\omega)
 \ten{G}_{me}(\vect{r},\vect{r}',\omega)
 \varepsilon(\vect{r}',\omega),\\
\label{eq30}
\ten{G}_{me}^\star(\vect{r},\vect{r}',\omega)
&=&-\varepsilon(\vect{r},\omega)
 \ten{G}_{em}(\vect{r},\vect{r}',\omega)
 \mu^{-1}(\vect{r}',\omega).
\end{eqnarray}

These laws immediately show that the discrete global duality
transformation $\varepsilon\!\leftrightarrow\!\mu$,
$\alpha\!\leftrightarrow\!\beta/c^2$ leaves ground-state dispersion
forces on stationary objects (and associated potentials) invariant,
where the individual electric and magnetic components~(\ref{eq20}),
(\ref{eq23}) and (\ref{eq26}) transform into one another according to
$\vect{F}_e\!\leftrightarrow\!\vect{F}_m$, 
$U_e\!\leftrightarrow\!U_m$ and $U_{ee}\!\leftrightarrow\!U_{mm}$, 
$U_{em}\!\leftrightarrow\!U_{me}$, respectively. Due to the factors
$\varepsilon$ and $\mu^{-1}$ appearing in
Eqs.~(\ref{eq27})--(\ref{eq30}), this only holds for forces on atoms
and bodies which are situated in free space. 

In order to extend duality invariance to atoms which are embedded in
a medium, local--field corrections need to be taken into account.
Using the real-cavity model, one can show that local field effects
give rise to correction factors\cite{0739}
\begin{equation}
\label{eq31}
c_e(\omega)=
\biggl[\frac{3\varepsilon(\omega)}
 {2\varepsilon(\omega)+1}\biggr]^2,\qquad
c_m(\omega)=
\biggl[\frac{3}{2\mu(\omega)+1}\biggr]^2,
\end{equation}
so the potentials~(\ref{eq23}) and (\ref{eq26}) generalize to
\begin{eqnarray}
\label{eq32}
U_\lambda(\vect{r}_A)
&=&\frac{\hbar}{2\pi\varepsilon_0}
 \int_0^\infty\dif\xi\,c_\lambda(\mi\xi)\alpha_\lambda(\mi\xi)
 \trace\ten{G}_{\lambda\lambda}^{(1)}(\vect{r}_A,\vect{r}_A,\mi\xi)
\quad(\lambda=e,m)\\
\label{eq33}
U_{\lambda\lambda'}(\vect{r}_A,\vect{r}_B)
&=&-\frac{\hbar}{2\pi\varepsilon_0^2}\int_0^\infty\dif\xi\,
 c_\lambda^A(\mi\xi)c_{\lambda'}^B(\mi\xi)
 \alpha_\lambda^A(\mi\xi)\alpha_{\lambda'}^B(\mi\xi)\nonumber\\
&&\times
 \trace\Bigl[\ten{G}_{\lambda\lambda'}(\vect{r}_A,\vect{r}_B,\mi\xi)
 \sprod\ten{G}_{\lambda'\lambda}(\vect{r}_B,\vect{r}_{\!A},\mi\xi)
 \Bigr],\quad(\lambda,\lambda'=e,m).
\end{eqnarray}
When applying a duality transformation, the factors $\varepsilon$ and
$\mu^{-1}$ arising from the transformation of the Green
tensor~(\ref{eq27})--(\ref{eq30}) combine with those contained in the
local-field correction factors~(\ref{eq31}) in such a way that the
corrected potentials~(\ref{eq32}) and (\ref{eq33}) transform into one
another according to $U_e\!\leftrightarrow\!U_m$,
$U_{ee}\!\leftrightarrow\!U_{mm}$ and
$U_{em}\!\leftrightarrow\!U_{me}$. When including local-field
corrections, the total ground-state dispersion potentials are hence
also duality invariant for embedded atoms.
%
%
\section{Summary}
\label{sec5}
We have studied the behavior of electromagnetic fields, response
functions and dispersion forces under duality transformations. In the
presence of media with a nontrivial impedance, the
$\operatorname{SO}(2)$ symmetry of duality rotations reduces to a
discrete $\mathbb{Z}_4$ symmetry. The duality transformations induced
by the generator of this group are displayed in Tab.~\ref{Tab1}. 
\begin{table}[t]
\tbl{Effect of the duality transformation with $r\!=1\!$,
$\theta=\pi/2$ on fields, response functions and dispersion forces.}
{
\begin{tabular}{cccc}
\hline
Dual partners&\multicolumn{3}{c}{Transformation}\\
\hline\hline
$\hat{\vect{E}}$, $\hat{\vect{H}}$:
&$\hat{\vect{E}}^\star=Z_0\hat{\vect{H}}$,&\hspace{1ex}&
$\hat{\vect{H}}^\star=-\hat{\vect{E}}/Z_0$\\
$\hat{\vect{D}}$, $\hat{\vect{B}}$:
&$\hat{\vect{D}}^\star=\hat{\vect{B}}Z_0$,&\hspace{1ex}&
$\hat{\vect{B}}^\star=-Z_0\hat{\vect{D}}$\\
$\hat{\vect{P}}$, $\hat{\vect{M}}$:
&$\hat{\vect{P}}^\star=\hat{\vect{M}}/c$,&\hspace{1ex}&
$\hat{\vect{M}}^\star=-c\hat{\vect{P}}$\\
$\hat{\vect{P}}_A$, $\hat{\vect{M}}_A$:
&$\hat{\vect{P}}_A^\star=\hat{\vect{M}}_A/c$,&\hspace{1ex}&
$\hat{\vect{M}}_A^\star=-c\hat{\vect{P}}_A$\\
$\hat{\vect{P}}_\mathrm{N}$, $\hat{\vect{M}}_\mathrm{N}$:
&$\hat{\vect{P}}_\mathrm{N}^\star
=\mu\hat{\vect{M}}_\mathrm{N}/c$,&\hspace{1ex}
&$\hat{\vect{M}}_\mathrm{N}^\star
=-c\hat{\vect{P}}_\mathrm{N}/\varepsilon$\\
$\hat{\vect{f}}_e$, $\hat{\vect{f}}_m$:
&$\hat{\vect{f}}_e^\star
=-\mi(\mu/|\mu|)\hat{\vect{f}}_m$,&\hspace{1ex}
&$\hat{\vect{f}}_m^\star
=-\mi(|\varepsilon|/\varepsilon)\hat{\vect{f}}_e$\\ 
\hline
$\varepsilon$, $\mu$:
&$\varepsilon^\star=\mu$,&\hspace{1ex}
&$\mu^\star=\varepsilon$\\
$\alpha$, $\beta$:
&$\alpha^\star=c^2\beta$,&\hspace{1ex}
&$\beta^\star=\alpha/c^2$\\
$\ten{G}_{ee}$, $\ten{G}_{mm}$:
&$\ten{G}^\star_{ee}=(1/\mu)\ten{G}_{mm}(1/\mu)+(1/\mu)\bm{\delta}$,
&\hspace{1ex}
&$\ten{G}^\star_{mm}=\varepsilon\ten{G}_{ee}\varepsilon
 -\varepsilon\bm{\delta}$\\
$\ten{G}_{em}$, $\ten{G}_{me}$:
&$\ten{G}^\star_{em}=-(1/\mu)\ten{G}_{me}\varepsilon$,
&\hspace{1ex}
&$\ten{G}^\star_{me}=-\varepsilon\ten{G}_{em}(1/\mu)$\\
\hline
$\vect{F}_e$, $\vect{F}_m$:
&$\vect{F}_e^\star=\vect{F}_m$,&\hspace{1ex}
&$\vect{F}_m^\star=\vect{F}_e$\\
$U_e$, $U_m$:
&$U_e^\star=U_m$,&\hspace{1ex}
&$U_m^\star=U_e$\\
$U_{ee}$, $U_{mm}$:
&$U_{ee}^\star=U_{mm}$,&\hspace{1ex}
&$U_{mm}^\star=U_{ee}$\\
$U_{em}$, $U_{me}$:
&$U_{em}^\star=U_{me}$,&\hspace{1ex}
&$U_{me}^\star=U_{em}$\\
\hline
\end{tabular}
\label{Tab1}
}
\end{table}
Note that the transformation needs to be applied four times to the
fields (first block) in order to return to the original state. On the
contrary, the transformation is self-inverse when acting on the
response functions of bodies, atoms and the electromagnetic field
(second block). We have shown that the electric and magnetic
components of dispersion forces on neutral bodies and atoms, which are
at rest and situated in free space, depend on these response functions
in such a way that they inherit a similar transformation behavior
(third block). As a consequence, the total dispersion forces are
duality invariant. As demonstrated, duality invariance can be extended
to atoms embedded in media provided that local-field corrections are
taken into account.

The duality invariance established in our work provides an important
consistency check for investigations of dispersion forces. It can
further serve as calculational tool: Once the force on an object in a
particular magnetoelectric environment is known, an expression for
the force in the dual arrangement can be generated by simply making
the replacements $\alpha\leftrightarrow\beta/c^2$,
$\varepsilon\leftrightarrow\mu$. 
%
%
\section*{Acknowledgments}
This work was supported by the Alexander von Humboldt Foundation and
the UK Engineering and Physical Sciences Research Council. H.S.\ would
like to thank the ministry of Science, Research, and Technology of
Iran for financial support. S.Y.B. is grateful to J.~Babington for
stimulating discussions.
%
%
\appendix
\section{Duality transformation of the Green tensor}
\label{appA}
To derive the transformed Green tensor $\ten{G}^\star$, which is a
solution to the Eq.~(\ref{eq18}) with $\varepsilon^\star\!=\!\mu$ and
$\mu^\star\!=\!\varepsilon$ instead of $\varepsilon$ and $\mu$, we
first note that the Maxwell equations~(\ref{eq1}) and (\ref{eq2})
together with the constitutive relations~(\ref{eq5}) are uniquely
solved by
\begin{eqnarray}
\label{a1}
\hat{\underline{\vect{E}}}(\vect{r},\omega)
&=&-\frac{1}{\varepsilon_0}\!\int\!\dif^3r'
 \ten{G}_{ee}(\vect{r},\vect{r}',\omega)
 \sprod\hat{\underline{\vect{P}}}_\mathrm{N}(\vect{r}',\omega)
 -Z_0\!\int\!\dif^3r'\ten{G}_{em}(\vect{r},\vect{r}',\omega)
 \sprod\hat{\underline{\vect{M}}}_\mathrm{N}(\vect{r}',\omega),
 \nonumber\\[-1ex] \\
\label{a2}
\hat{\underline{\vect{B}}}(\vect{r},\omega)
&=&-Z_0\!\int\!\dif^3r'
 \ten{G}_{me}(\vect{r},\vect{r}',\omega)
 \sprod\hat{\underline{\vect{P}}}_\mathrm{N}(\vect{r}',\omega)
 -\mu_0\!\int\!\dif^3r'
 \ten{G}_{mm}(\vect{r},\vect{r}',\omega)
 \sprod\hat{\underline{\vect{M}}}_\mathrm{N}(\vect{r}',\omega),
 \nonumber\\[-1ex] \\[-1ex]
\label{a3}
\hat{\underline{\vect{D}}}(\vect{r},\omega)
&=&-\frac{\varepsilon(\vect{r},\omega)}{c}\!
 \int\!\dif^3r'\ten{G}_{em}(\vect{r},\vect{r}',\omega)
 \sprod\hat{\underline{\vect{M}}}_\mathrm{N}(\vect{r}',\omega)
 \nonumber\\
&&-\int\!\dif^3r'\biggl[\varepsilon(\vect{r},\omega)
\ten{G}_{ee}(\vect{r},\vect{r}',\omega)
 -\bm{\delta}(\vect{r}\!-\!\vect{r}')\biggr]
 \sprod\hat{\underline{\vect{P}}}_\mathrm{N}(\vect{r}',\omega),
 \\
\label{a4}
\hat{\underline{\vect{H}}}(\vect{r},\omega)
&=&-\frac{c}{\mu(\vect{r},\omega)}\!\int\!\dif^3r'
 \ten{G}_{me}(\vect{r},\vect{r}',\omega)
 \sprod\hat{\underline{\vect{P}}}_\mathrm{N}(\vect{r}',\omega)
 \nonumber\\
&&-\int\!\dif^3r'\biggl[\frac{\ten{G}_{mm}(\vect{r},\vect{r}',\omega)}
 {\mu(\vect{r},\omega)}
 +\bm{\delta}(\vect{r}\!-\!\vect{r}')\biggr]
 \sprod\hat{\underline{\vect{M}}}_\mathrm{N}(\vect{r}',\omega).
\end{eqnarray}
Applying the duality transformation to Eqs.~(\ref{a1}) and (\ref{a2})
with the aid of the transformation laws established in
Sec.~\ref{sec2} and using Eqs.~(\ref{a3}) and (\ref{a4}), the unknown
quantities $\ten{G}_{\lambda\lambda'}^\ast$ on the rhs of the
transformed equations can be related to the untransformed ones
$\ten{G}_{\lambda\lambda'}$ appearing on the lhs, and one obtains
Eqs.~(\ref{eq27})--(\ref{eq30}).
%

%
\end{document}